\documentclass{iopart}
\usepackage{graphicx}  
\usepackage{latexsym}  

\jl{6}        
\eqnobysec    

\def\beq{\begin{equation}}
\def\eeq{\end{equation}}
\def\rmd{{\rm d}}
\def\rightcontract{\mathop{\hbox{\vrule width0.5pt height6pt%
  \vrule height0.5pt width6pt}}}

\begin{document}

\title[Spinning test particles and clock effect in Kerr spacetime]
{Spinning test particles and clock effect in Kerr spacetime}

\author{
Donato Bini$^* {}^\S{}^\P$,
Fernando de Felice$^\dagger$ 
and 
Andrea Geralico$^\ddag {}^\S$}
\address{
  ${}^*$\
Istituto per le Applicazioni del Calcolo ``M. Picone'', CNR I-00161 Rome, Italy
}
\address{
  ${}^\S$\
  International Center for Relativistic Astrophysics,
  University of Rome, I-00185 Rome, Italy
}
\address{
${}^\P$
  INFN - Sezione di Firenze, Polo Scientifico, Via Sansone 1, 
  I-50019, Sesto Fiorentino (FI), Italy 
}
\address{
${}^\dagger$\
Dipartimento di Fisica, Universit\`a di Padova, and INFN, Sezione di Padova, Via Marzolo 8,  I-35131 Padova, Italy}
\address{
  ${}^\ddag$\
  Dipartimento di Fisica, Universit\`a di Lecce, and INFN - Sezione di Lecce,
  Via Arnesano, CP 193, I-73100 Lecce, Italy}

\begin{abstract}
We study the  motion of spinning test particles in Kerr spacetime using the Mathisson-Papa\-pe\-trou equations; we impose different supplementary conditions among the well known Cori\-nal\-de\-si-Papapetrou, Pirani and Tulczyjew's and analyze  their physical implications in order to decide which is the most natural to use.
We find that if the particle's center of mass world line, namely the one chosen for the multipole reduction,  is  a spatially circular orbit (sustained by the tidal forces due to the spin) then  the generalized momentum $P$ of the test particle  is also tangent to a spatially circular orbit intersecting the center of mass line at a point. There exists one such orbit for each point of the center of mass line where they intersect; although fictitious, these orbits are essential 
to define the properties of the spinning particle along its physical motion.  
In the  small spin limit, the  particle's orbit is almost a geodesic and
the difference of its angular velocity with respect to the geodesic value can be of arbitrary sign, corresponding to the  spin-up and spin-down possible alignment along the $z$-axis. We also find that the choice of the supplementary conditions leads to clock effects of substantially different magnitude. In fact, for 
 co-rotating and counter-rotating  particles having the same spin magnitude and orientation, the gravitomagnetic clock effect induced by the background metric can be magnified or inhibited and even suppressed 
by the contribution of the individual particle's spin.  Quite surprisingly this contribution can be itself made vanishing leading to a clock effect undistiguishable from that of non spinning particles.
The results of our analysis can be observationally tested.
\end{abstract}

\pacno{04.20.Cv}

\section{Introduction}

The motion of a spinning test particle is described by 
the Mathisson-Papapetrou \cite{math37,papa51} equations
\begin{eqnarray}
\label{papcoreqs1}
\frac{DP^{\mu}}{\rmd \tau_U}&=&-\frac12R^{\mu}{}_{\nu\alpha\beta}U^{\nu}S^{\alpha\beta}\equiv F^{\rm (spin)}{}^{\mu}\\
\label{papcoreqs2}
\frac{DS^{\mu\nu}}{\rmd \tau_U}&=&P^{\mu}U^{\nu}-P^{\nu}U^{\mu}\ ;
\end{eqnarray}
here $P$ is the total momentum of the particle,  $S=1/2 S_{\alpha\beta} \, \omega^\alpha \wedge \omega^\beta $ is the antisymmetric spin tensor, where $\{\omega^\alpha\}$ is a suitable tetrad frame, and $U$ is the timelike unit tangent vector of the particle's \lq\lq center line'' used to make the multipole reduction.
The test character of the particle under consideration refers either to its mass-energy content or to the spin, since both quantities should not be 
large enough to significatively perturbe the background metric. In what follows, with the magnitude of the spin of the particle, with the mass and  with a natural lenghtscale associated to the gravitational background we will introduce an adimensional quantity as a smallness indicator, which we retain to the first order only so that the test character of the particle  be fully satisfied. 
Moreover, it is well known that, in order to close the above set of differential equations, one must add supplementary conditions (SC) for which standard choices in the literature are
\begin{itemize}
\item[1.] 
Papapetrou-Corinaldesi \cite{cori51} conditions (PC): $S^{t\nu}=0$,
\item[2.] 
Pirani \cite{pir56} conditions (P): $S^{\mu\nu}U_\nu=0$, 
\item[3.] 
Tulczyjew \cite{tulc59} conditions (T): $S^{\mu\nu}P_\nu=0$.
\end{itemize}
Following the work of Tod, de Felice and Calvani \cite{tod77}
we will discuss here the behaviour of spinning test particles in spatially circular motion on the equatorial plane of the Kerr spacetime, with $U$ (and consequently also $P$, as it will be shown in the following section)  aligned along a circular orbit.
For certain choices of spin alignement the \lq\lq spin force", which  couples the spin of the particle  to the background curvature, is able to 
maintain the orbit; in the absence of spin, instead, only geodesic motion  is allowed, unless other external forces were applied.
A recent work on spinning  particles in the Kerr spacetime is due Hartl \cite{hartl1,hartl2} and to this work we refer also for an updated and complete bibliography.

In General Relativity the time rate of a clock with respect to a given observer depends on the background geometry and the specific properties of their spacetime motion. The asymmetry of time reading of two clocks when they intersect after having moved from the same spacetime point along trajectories with opposite azimuthal angular momentum is termed {\it clock effect}. In this paper we shall deduce the clock effect for spinning particles moving on circular orbits in Kerr spacetime depending on the supplementary condition one adds to the equations of motion.
Some of the results of this paper were found in \cite{abramcalv} concerning the properties of the circular orbits and more recently in \cite{faruq} concerning the clock effect; either papers however do not consider the dependence of particle dynamics and clock effect on the supplementary conditions and the spin relative orientations.  

The structure of the paper is as follows. In Section 2 we analyse the general properties of the circular orbits followed by spinning particles in Kerr spacetime. Then in subsections 2.1 to 2.3 we specify the supplementary conditions and identify those leading to unphysical conditions and therefore to be disregarded. In Section 3 we analyse the clock effects under the {\it contraints} induced by different supplementary conditions and see, consistently with the results of the previous section, which one can be experimentally justified.
Finally in Section 4 we draw our conclusions.

Hereafter we shall use geometrized units; Greek indeces run from $0$ to $3$ and Latin indices run from $1$ to $3$. 
The metric signature is chosen as $+2$.

\section{Dynamics of spinning particles: circular orbits in Kerr spacetime}

The Kerr metric  in standard Boyer-Lindquist coordinates is given by
\begin{eqnarray}
\rmd s^2 &=& -\left(1-\frac{2Mr}{\Sigma}\right)\rmd t^2 -\frac{4aMr}{\Sigma}\sin^2\theta\rmd t\rmd\phi+ \frac{\Sigma}{\Delta}\rmd r^2 +\Sigma\rmd \theta^2\nonumber\\
&&+\frac{(r^2+a^2)^2-\Delta a^2\sin^2\theta}{\Sigma}\sin^2 \theta \rmd \phi^2\ ,
\end{eqnarray}
where $\Delta=r^2-2Mr+a^2$ and $\Sigma=r^2+a^2\cos^2\theta$; here $a$ and $M$ are the specific angular momentum and total mass of the spacetime solution. Let us introduce the ZAMO family of fiducial observers, with four velocity
\beq
\label{n}
n=N^{-1}(\partial_t-N^{\phi}\partial_\phi);
\eeq
here $N=(-g^{tt})^{-1/2}$ and $N^{\phi}=g_{t\phi}/g_{\phi\phi}$ are the lapse and shift functions respectively. A suitable orthonormal frame adapted to  ZAMOs is given by
\beq
e_{\hat t}=n , \,\quad
e_{\hat r}=\frac1{\sqrt{g_{rr}}}\partial_r, \,\quad
e_{\hat \theta}=\frac1{\sqrt{g_{\theta \theta }}}\partial_\theta, \,\quad
e_{\hat \phi}=\frac1{\sqrt{g_{\phi \phi }}}\partial_\phi ,
\eeq
with dual 
\beq\fl\quad
\omega^{{\hat t}}=N\rmd t\ , \quad \omega^{{\hat r}}=\sqrt{g_{rr}}\rmd r\ , \quad 
\omega^{{\hat \theta}}= \sqrt{g_{\theta \theta }} \rmd \theta\ , \quad
\omega^{{\hat \phi}}=\sqrt{g_{\phi \phi }}(\rmd \phi+N^{\phi}\rmd t)\ .
\eeq
In terms of (\ref{n}) the line element can be expressed in the form
\beq
\rmd s^2 = -N^2\rmd t^2 +g_{\phi \phi }(\rmd \phi+N^{\phi}\rmd t)^2 + g_{rr}\rmd r^2 +g_{\theta \theta}\rmd \theta^2\ . 
\eeq

The four-velocity $U$ of uniformly rotating circular orbits
can be parametrized either by the (constant) angular velocity with respect to infinity $\zeta$ or, equivalently, by the (constant) linear velocity  $\nu$ with respect to ZAMOs 
\beq
\label{orbita}
U=\Gamma [\partial_t +\zeta \partial_\phi ]=\gamma [e_{\hat t} +\nu e_{\hat \phi}], \qquad \gamma=(1-\nu^2)^{-1/2}\ ,
\eeq
where $\Gamma$ is a normalization factor which assures that $U_\alpha U^\alpha =-1$  given by:
\beq
\Gamma =\left[ N^2-g_{\phi\phi}(\zeta+N^{\phi})^2 \right]^{-1/2}=\frac{\gamma}{N}
\eeq
 and
\beq
\zeta=-N^{\phi}+\frac{N}{\sqrt{g_{\phi\phi}}} \nu .
\eeq
We limit our analysis to  the equatorial plane ($\theta=\pi/2$) of the Kerr solution; as a convention, the physical (orthonormal) component along $-\partial_\theta$, perpendicular to the equatorial plane will be referred to as along the positive $z$-axis and will be indicated by $\hat z$, when necessary. The spacetime trajectory described by (\ref{orbita}) will be termed $U$-orbit.

On the equatorial plane of the Kerr solution there exists a large variety of special circular orbits  \cite{bjdf,idcf1, idcf2,bjm}; particular interest is devoted to the co-rotating $(+)$ and counter-rotating $(-)$ timelike circular geodesics whose angular and linear velocities are respetively
\beq\fl\quad
\zeta_{({\rm geo})\, \pm}\equiv\zeta_{\pm}=\left[a\pm (M/r^3)^{1/2}\right]^{-1}\ , \quad 
\nu_{({\rm geo})\, \pm}\equiv \nu_\pm =\frac{a^2\mp2a\sqrt{Mr}+r^2}{\sqrt{\Delta}(a\pm r\sqrt{r/M})}\ .
\eeq 
Other special orbits correspond to the \lq\lq geodesic meeting point observers'' first defined in \cite{idcf2}, with
\beq
\nu_{\rm (gmp)}=\frac{
\nu_{+}+\nu_{-}}{2}=-\frac{aM(3r^2+a^2)}{\sqrt{\Delta}(r^3-a^2M)}\ ,
\eeq
and those characterized by
\beq
\nu_{\mathcal{Z}{\rm (nmp)}}=\frac{2}{\nu_{+}^{-1}+\nu_{-}^{-1}}=\frac{(r^2+a^2)^2-4a^2Mr}{a\sqrt{\Delta}(3r^2+a^2)}\ ,
\eeq
both of them playing a role in the study of parallel transport of vectors along circular orbits (see \cite{bjm} for the properties of the map $\mathcal{Z}(\nu)$ and those of the \lq\lq null meeting point observers'' or \lq\lq ZAMOs'' as termed in that work due to the fact that their world lines contain the meeting points of co/counter rotating photons in parallel to the \lq\lq geodesic meeting point observers'', whose world lines contain the meeting points of co/counter rotating geodesics).

It is convenient to introduce the Lie relative curvature of each orbit \cite{idcf2}
\beq
k_{\rm (lie)}=-\partial_{\hat r} \ln \sqrt{g_{\phi\phi}}=-\frac{(r^3-a^2M)\sqrt{\Delta}}{r^2(r^3+a^2r+2a^2M)}\ ,
\eeq
as well as
a Frenet-Serret (FS) intrinsic frame along $U$ \cite{iyer-vish}, both well established in the literature. 
It is well known that any circular orbit on the equatorial plane of the Kerr spacetime has zero second torsion $\tau_2$, while the geodesic curvature $\kappa$ and the first torsion $\tau_1$ are simply related by
\beq
\tau_1= -\frac{1}{2\gamma^2} \frac{\rmd \kappa}{\rmd \nu}\ ,
\eeq
so that 
\begin{eqnarray}
\label{ketau1}
\kappa &=&k_{\rm (lie)}\gamma^2 (\nu-\nu_+)(\nu-\nu_-)\ , \nonumber \\
\tau_1&=& k_{\rm (lie)}\nu_{\rm (gmp)}\gamma^2 (\nu-\nu_{{\rm (crit)}+})(\nu-\nu_{{\rm (crit)}-})\ ,
\end{eqnarray}
where
\begin{eqnarray}
\fl\quad\nu_{{\rm (crit)}\pm}&=&\frac{\gamma_- \nu_- \mp \gamma_+ \nu_+}{\gamma_- \mp \gamma_+} \nonumber \\
\fl\quad&=& -\frac1{2Ma(3r^2+a^2)\sqrt{\Delta}}\Big[-2a^2M(a^2-3Mr)+r^2(r^2+a^2)(r-3M)\nonumber\\
\fl\quad &&\pm(r^3+a^2r+2a^2M)\sqrt{r}\sqrt{r(r-3M)^2-4a^2M}\Big]\
\end{eqnarray}
identify the so called \lq\lq extremely accelerated observers'' first introduced in \cite{fdfacc}.
The FS frame along $U$  is then given by
\begin{eqnarray}\fl\quad
E_0\equiv U=\gamma [e_{\hat t}+\nu e_{\hat \phi}],\quad E_1=e_{\hat r}, \quad E_2\equiv E_{\hat \phi}=\gamma [\nu n+e_{\hat \phi}],\quad  E_3=e_{\hat z}\ , 
\end{eqnarray}
satisfying the following system of evolution equations
\begin{eqnarray}
\label{FSeqs}
\frac{DE_0}{d\tau_U}&=&\kappa E_1\ ,\qquad\quad\,\,\,  
\frac{DE_1}{d\tau_U}=\kappa E_0+\tau_1 E_2\ ,\nonumber \\
 \nonumber \\
\frac{DE_2}{d\tau_U}&=&-\tau_1E_1\ , \qquad 
\frac{DE_3}{d\tau_U}=0\ .
\end{eqnarray}

To study circularly rotating spinning test particles let us consider first the evolution equation of the spin tensor (\ref{papcoreqs2}), 
assuming the frame components of the spin tensor as constant along the orbit. 
Following the analysis made in \cite{bdfg1}
the total four-momentum can be written as
\begin{equation}
\label{Ps}
P^{\mu}=-(U\cdot P)U^\mu -U_\nu \frac{DS^{\mu\nu}}{\rmd \tau_U}\equiv
mU^\mu +P_s^\mu\ ,
\end{equation}
where $P_s=U \rightcontract DS/{\rmd \tau_U}$ ($\rightcontract$ denoting the right contraction operation between tensors in index-free form) and $m$ is the mass  the particle would have in the rest space of $U$ if it were not spinning; we shall denote it as {\it bare mass}. 
From (\ref{Ps}), Eq.~(\ref{papcoreqs2}) is also equivalent to
\begin{equation}
\label{proiet}
P(U)^\mu_{\alpha}P(U)^\nu_{\beta}\frac{DS^{\alpha\beta}}{\rmd \tau_U}=0\ ,
\end{equation}
where $P(U)^\mu_\alpha=\delta^\mu_\alpha+U^\mu U_\alpha$  projects in  the  rest space of $U$; it implies
\begin{equation}
\label{spinconds}
S_{\hat t\hat \phi}=0\ , 
\quad S_{\hat r\hat \theta}=0\ , 
\quad S_{\hat t\hat \theta}+ S_{\hat \phi \hat \theta}\frac{\nu-\nu_{\rm (gmp)}}{\nu_{\rm (gmp)}(\nu-\nu_{\mathcal{Z}({\rm nmp})})}=0 \ .
\end{equation}
It is clear from (\ref{Ps}) that $P_s$ is orthogonal to $U$; moreover it turns out to be also aligned with $E_{\hat \phi}$  
\begin{equation}
\label{ps}
P_s=m_s E_{\hat \phi}\ ,
\end{equation}
where $m_s\equiv||P_s||$ is given by
\begin{eqnarray}
\label{msdef}
m_s=-\gamma k_{\rm (lie)}
\left[
S_{\hat t\hat r}(\nu-\nu_{\rm (gmp)})+S_{\hat r\hat \phi}\nu_{\rm (gmp)}(\nu-\nu_{\mathcal{Z}({\rm nmp})})\right]\ .
\end{eqnarray}
From (\ref{Ps}) and (\ref{ps}) the total four-momentum $P$  can be written in the form $P=\mu \, U_p$, with
\begin{equation}
\label{Ptot}
U_p=\gamma_p\, [e_{\hat t}+\nu_p e_{\hat \phi}]\ , \quad \nu_p=\frac{\nu+m_s/m}{1+\nu m_s/m}\ ,\quad \mu=\frac{\gamma}{\gamma_p}(m+\nu m_s)\ 
\end{equation}
and $\gamma_p=(1-\nu_p^2)^{-1/2}$.

Since $U_p$ is a unit vector, the quantity $\mu$ can be interpreted as the total mass of the particle in the rest-frame of $U_p$.
We see from (\ref{Ptot}) that the total four-momentum $P$ is parallel to the unit tangent of a spatially circular orbit that we shall denote as $U_p$-orbit. The latter intersects the $U$-orbit only at one point where only it makes sense to compare the vectors $U$ and $U_p$ and the physical quantities related to them. It is clear that there exists one $U_p$-orbit for each point of the $U$-orbit where the two intersect hence, along the $U$-orbit, we can only compare {\it at} the point of intersection the quantities defined in a frame adapted to $U$ with those defined in a frame adapted to $U_p$. 

Let us now consider  the equation of motion (\ref{papcoreqs1}). 
After some algebra the spin-force turns out to be equal to  
\begin{eqnarray}
\label{fspin}
\fl\quad F^{\rm (spin)}&=&\frac{\gamma }{r^4}\frac{M}{r(r^2+a^2)+2a^2M}\Big\{[r^2(2r^2+5a^2)+a^2(3a^2-2Mr)\nonumber\\
\fl\quad &&-3a(r^2+a^2)\sqrt{\Delta}\nu] S_{\hat t\hat r}+\{[r^2(r^2+4a^2)+a^2(3a^2-4Mr)]\nu\nonumber\\
\fl\quad &&-3a(r^2+a^2)\sqrt{\Delta}\}S_{\hat r\hat \phi}\}\Big\}e_{\hat r}\nonumber\\
\fl\quad &&+\frac{\gamma}{r^3}\frac{S_{\hat \theta\hat \phi }}{(r^2+a^2)^2-4a^2Mr-a(3r^2+a^2)\sqrt{\Delta }\nu}\Big\{-\nu[3a^2r(a^2-2Mr)\nonumber\\
\fl\quad &&+r^3(r^2+4a^2)-2Ma^4]+a\sqrt{\Delta}\{-4Ma^2+\nu^2[3r(r^2+a^2)\nonumber\\
\fl\quad &&+2a^2M]\}\Big\} e_{\hat \theta}\ ,
\end{eqnarray}
while the term on the left hand side of Eq.~(\ref{papcoreqs1}) can be written as
\beq
\label{dpdtau}
\frac{DP}{\rmd \tau_U}=m a(U)+m_s \frac{DE_{\hat \phi}}{\rmd \tau_U}\ ,
\eeq
where
\beq
a(U)=\kappa e_{\hat r}\ , \qquad \frac{DE_{\hat \phi}}{\rmd \tau_U}=-\tau_1 e_{\hat r}=\frac{1}{2\gamma^2} \frac{\rmd \kappa}{\rmd \nu} e_{\hat r}\ ,
\eeq
$\mu, m, m_s$ being constant along the $U$-orbit.
We note that the term $DE_{\hat \phi}/\rmd \tau_U$ represents a sort of centrifugal force (modulo factors) measured by the \lq\lq observer" $U$  in its rest frame: this has been studied in detail in \cite{bjdf,idcf1,idcf2}. 
The spin-rotation coupling exhibited by the last term in Eq.~(\ref{dpdtau})
is the classical manifestation of the effect that was first predicted by Mashhoon \cite{mash88}.

Since $DP/{\rmd \tau_U}$ is directed radially,  Eq.~(\ref{fspin}) requires that $S_{\hat \theta\hat \phi}=0$ (and therefore, from (\ref{spinconds}), $S_{\hat t\hat \theta}=0$); Eq.~(\ref{papcoreqs1}) then reduces to
\begin{eqnarray}
\label{eqmoto}
m\kappa -m_s\tau_1-F^{\rm (spin)}_{\hat r}=0\ . 
\end{eqnarray}

Summarizing, from the equations of motions (\ref{papcoreqs1}) and (\ref{papcoreqs2}) we deduce that the spin tensor is completely determined by  two components, namely $S_{\hat t\hat r}$ and $S_{\hat t\hat \phi}$,
so that
\begin{equation}
S=\omega^{\hat r}\wedge [S_{\hat r\hat t}\omega^{\hat t}+S_{\hat r\hat \phi}\omega^{\hat \phi}]\ .
\end{equation}
From the relations 
\begin{eqnarray}
\omega^{\hat t}&=&\gamma [-U^{\flat}+\nu \Omega^{\hat \phi}]\ , \nonumber\\
\omega^{\hat \phi}&=& \gamma [-\nu U^{\flat} +\Omega^{\hat \phi}], \qquad \Omega^{\hat \phi}=[E_{\hat \phi}]^{\flat}\ ,
\end{eqnarray}
where $X^{\flat}$ denotes the 1-form associated to a vector $X$,
one has the useful relation
\begin{equation}
S=\gamma \left[(S_{\hat r\hat t} + \nu S_{\hat r\hat \phi} )  U^{\flat}\wedge \omega^{\hat r} +(\nu S_{\hat r\hat t} + S_{\hat r\hat \phi} )\omega^{\hat r}\wedge \Omega^{\hat \phi} \right]\ .
\end{equation}
Since the components of $S$ are constant along $U$, then from the FS formalism one finds
\beq
\frac{DS}{\rmd \tau_U}=\gamma [(\tau_1+\kappa \nu )S_{\hat r\hat t}+(\nu \tau_1+\kappa )S_{\hat r\hat \phi}] U^{\flat} \wedge \Omega^{\hat \phi}\ ,
\eeq
or from Eq.~(\ref{papcoreqs2})
\beq
P_s=-\gamma [(\tau_1+\kappa \nu )S_{\hat r\hat t}+(\nu \tau_1+\kappa )S_{\hat r\hat \phi}] \Omega^{\hat \phi}\equiv m_s \Omega^{\hat \phi}\ ,
\eeq
where we introduce the quadratic invariant 
\beq
s^2=\frac12 S_{\mu\nu}S^{\mu\nu}=-S_{\hat r\hat t }^2+S_{\hat r \hat \phi}^2\ .
\eeq
To discuss the physical properties of the particle motion one needs to supplement Eq.~(\ref{eqmoto}) with ($3$ at most) further conditions because 
Eq.~(\ref{papcoreqs2}) is identically satisfied if the only nonzero components of $S$ are $S_{\hat r\hat t }$ and $S_{\hat r \hat \phi}$. 
All the standard approaches existing in the literature can be summarized by the following choice:
\beq
\label{SCgen}
(S_{\hat r\hat t}, S_{\hat r\hat \phi})=s(-\tilde \gamma \tilde \nu , \tilde \gamma)\ .
\eeq
Relations (\ref{SCgen})  are equivalent to $S_{\alpha\beta} \tilde U{}^\beta=0$ for some timelike vector $\tilde U$.
When $\tilde U=1/\sqrt{-g_{tt}}\partial_t$, i.e. $\tilde \nu=-\frac{2aM}{r\sqrt{\Delta }}$, we have the Corinaldesi-Papapetrou supplementary conditions; 
when $\tilde U=U$, i.e. $\tilde \nu=\nu$, we have the Pirani's conditions, and when $\tilde U=U_p$, i.e. $\tilde \nu=\nu_p$, we have the Tulczyjew's conditions. 
Before discussing each of these conditions separately let us summarize the results.
The quantity $m_s$ in general is given by
\beq
\label{msdef2}
m_s=-s\tilde \gamma \gamma [-\tilde \nu (\tau_1+\kappa \nu )+(\nu \tau_1+\kappa )]\ ,
\eeq
and, once inserted in the equation of motion, it gives
\begin{eqnarray}
\label{eqmoto_1}
m\kappa +s\tilde \gamma \gamma [-\tilde \nu (\tau_1+\kappa \nu )+(\nu \tau_1+\kappa )]
\tau_1-F^{\rm (spin)}_{\hat r}=0\ , 
\end{eqnarray}
where $F^{\rm (spin)}_{\hat r}$ is given by
\beq
F^{\rm (spin)}_{\hat r}= s\gamma \tilde \gamma [A \nu \tilde \nu + B\nu +C\tilde \nu +A]\ ,
\eeq
with
\begin{eqnarray}
A&=&-\frac{3Ma(r^2+a^2)\sqrt{\Delta}}{r^4(r^3+a^2r+2a^2M)}\nonumber\\
B&=&\frac{M}{r^4}\frac{r^4+3a^4+4a^2r(r-M)}{r^3+a^2r+2a^2M}\nonumber\\
C&=&B+\frac{M}{r^3}\ .
\end{eqnarray}
Thus one can solve Eq.~(\ref{eqmoto_1}) for the quantity ${\hat s}=\pm |{\hat s}|=\pm |s|/(mM)$, 
which denotes the signed magnitude of the spin per unit (bare) mass $m$ of the test particle and $M$ of the black hole, obtaining
\beq\fl\quad 
\label{ssolgen}
\hat s=-\frac{\kappa}{M\gamma \tilde \gamma \left\{
[-\tilde \nu (\tau_1+\kappa \nu )+(\nu \tau_1+\kappa )]\tau_1-(A \nu \tilde \nu + B\nu +C\tilde \nu +A)
\right\}}\ ,
\eeq
with $\kappa$  and $\tau_1$ given by (\ref{ketau1}).

\subsection{The Corinaldesi-Papapetrou (CP) supplementary conditions}

\begin{figure} 
\typeout{*** EPS figure 1}
\begin{center}
\includegraphics[scale=0.35]{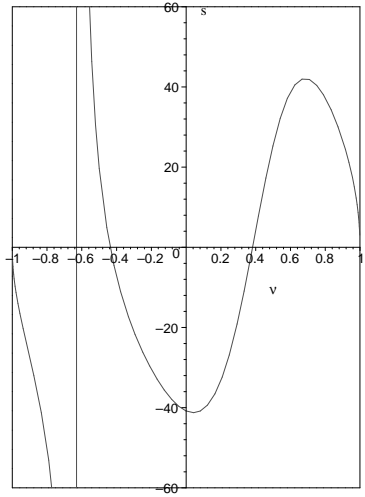}
\end{center}
\caption{In the case of CP supplementary conditions, the spin parameter ${\hat s}$ is plotted as a function 
of the linear velocity $\nu$, for $M=1$, $a=0.5$ and $r=8$.
From (\ref{ssolgen}) we have that ${\hat s}$ vanishes for $\kappa=0$, or $\nu=\nu_{\pm}\approx-0.436, 0.382$
as well as for $\nu=\pm 1$. ${\hat s}$ diverges instead at $\nu\approx -0.631$, showing that this value of the velocity
is not allowed. Actually not only this value is not allowed but all those corresponding to $\hat s \gg 1$ should be considered of little physical significance because, as stated in the introduction, the spinning particle would be no more a test particle. An analogous remark also applies to the next figures 3 and 4.
}  
\label{fig:1}
\end{figure}

\begin{figure} 
\typeout{*** EPS figure 2}
\begin{center}
$
\begin{array}{cc}
\includegraphics[scale=0.35]{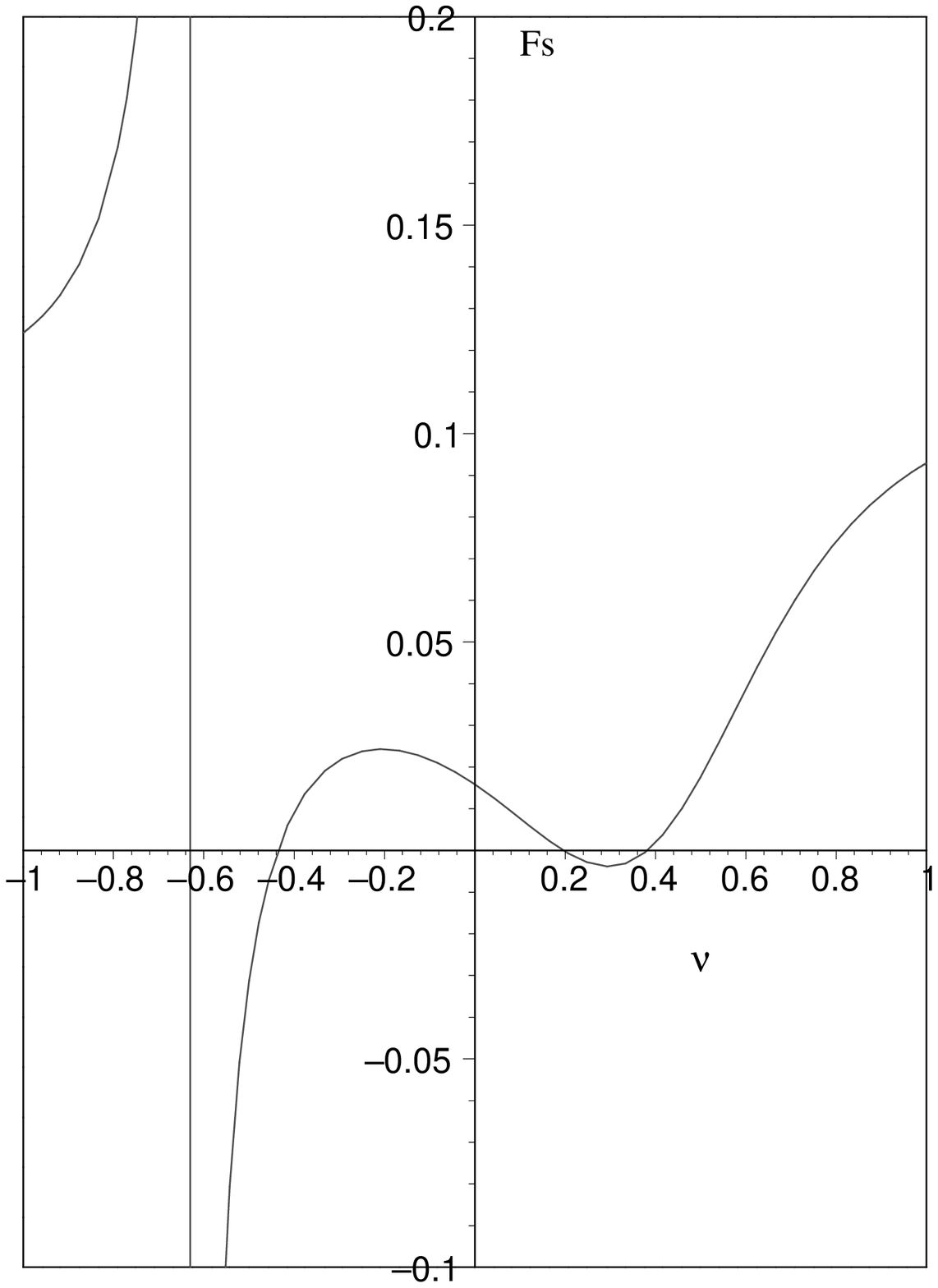}&\quad
\includegraphics[scale=0.35]{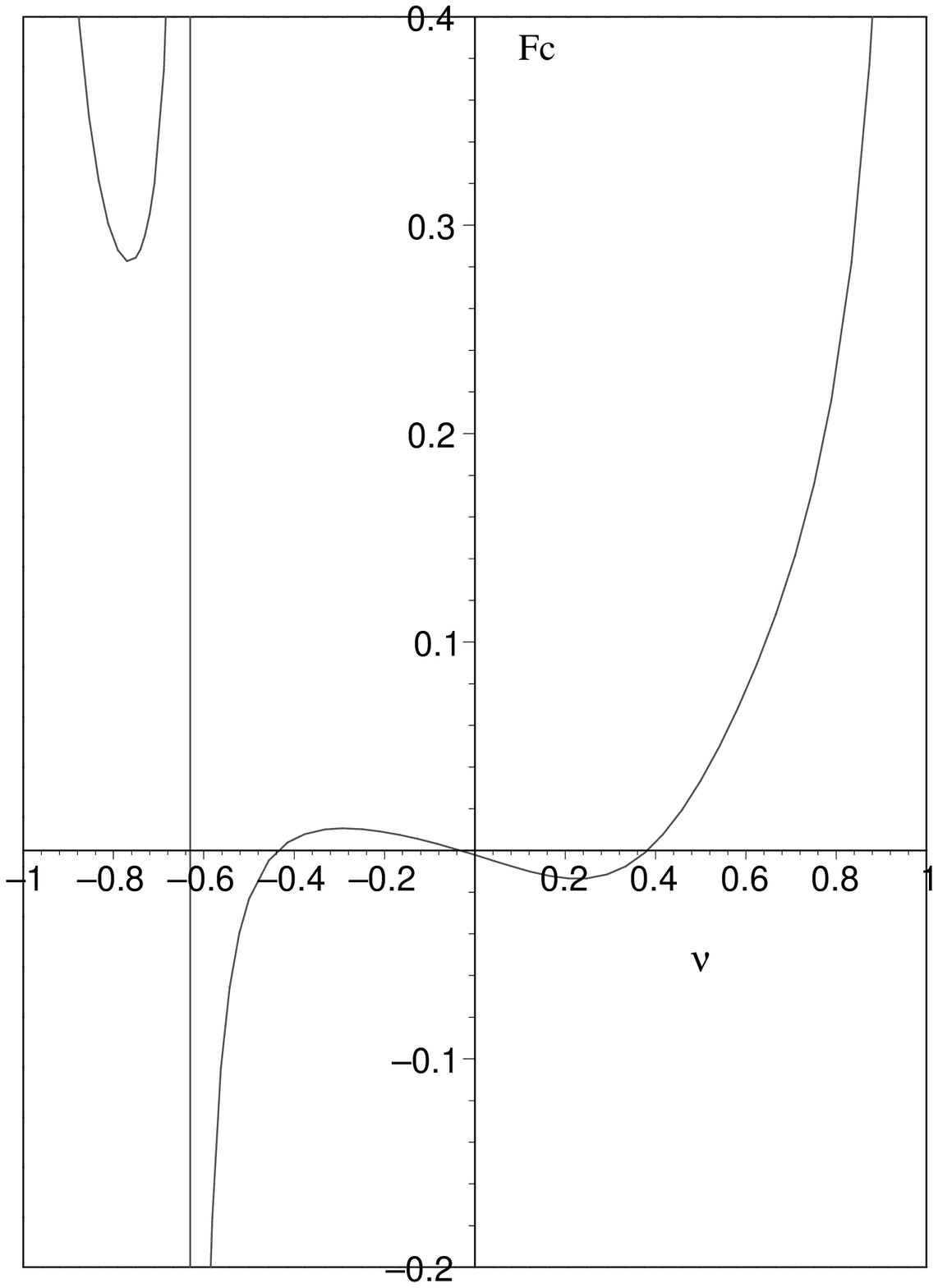}\\[.4cm]
\quad\mbox{(a)}\quad &\quad \mbox{(b)}
\end{array}
$\\[.6cm]
\includegraphics[scale=0.35]{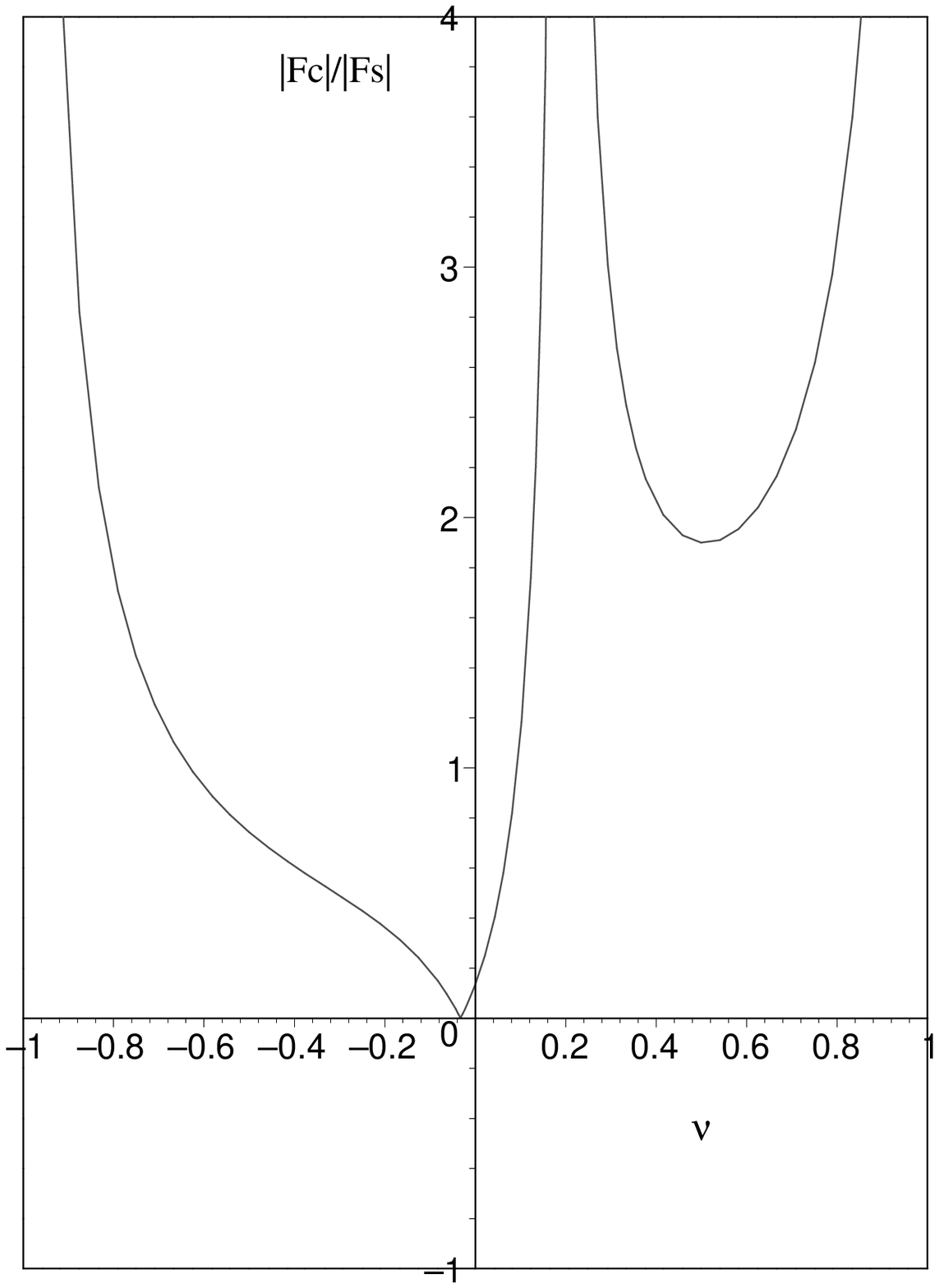}\\[.4cm]
\quad\mbox{(c)}
\end{center}
\caption{
In the case of CP supplementary conditions, the behaviours of the spin  and centrifugal forces and their ratio as functions of $\nu$ are shown in 
Fig. (a) - (c) respectively for $M=1$, 
$a=0.5$ and $r=8$. Both the force due to the spin and the centrifugal forces diverge at $\nu=-0.631$ because here $\hat s$ diverges at this value. Their ratio instead is finite because it is no longer dependent on $\hat s$.}  
\label{fig:2}
\end{figure}

The CP supplementary conditions are given by (\ref{SCgen}) with $\tilde \nu=-\frac{2aM}{r\sqrt{\Delta }}$.  
By substituting it into Eq.~(\ref{ssolgen}), we obtain the relevant relation between  ${\hat s}$ and $\nu$ which is plotted
in Fig. \ref{fig:1}, for a fixed value of the radial coordinate $r$ and  of the 
 parameters $a$ and $M$. 
Comparisons between the centrifugal 
($Fc=-m_s DE_{\hat \phi}/\rmd \tau_U$) and spin ($Fs$ in the plot) forces are shown in Fig. \ref{fig:2}.
It is worth noting that the divergence of the spin $\hat s$ when $\nu=0$  in the Schwarzschild case (see \cite{bdfg1})  is here found at a negative value of $\nu$ (counter-rotating circular orbit) due to the asymmetry  induced  by the spacetime 
rotation. While we judge the divergence of $\hat s$ a symptom of the inadequacy of the CP conditions at large spins, most interesting is the vanishing of the spin force not only when the orbit is a geodesic, but also when $\nu$  equals  a positive non geodesic value (see Fig. \ref{fig:2} (a)).
This effect arises also in the cases we shall consider next.
Clearly this seems to be the result of a compensation among the particle spin, the spacetime rotation and the orbital angular momentum.


In the case of small spins, namely if  ${\hat s}\ll1$, we have to first order in $\hat s$
\begin{eqnarray}
\label{solCPexpnu}
\nu&=& \nu_{\pm}+{\mathcal N}^{(CP)}{\hat s}+O({\hat s}^2)\nonumber\\
{\mathcal N}^{(CP)}&=&-\frac32\frac{aM(r^3+a^2r+2a^2M) [\sqrt{Mr}(r-3M)\pm2aM]^{1/2} }{r\sqrt{\Delta } \sqrt{r-2M}(Mr)^{3/4} (a\pm r\sqrt{r/M})^2}\ .
\end{eqnarray}
The corresponding angular velocity $\zeta$ and its reciprocal are
\begin{eqnarray}
\label{zetaCP}
\zeta&=& \zeta_{\pm} \left[1-\frac32\frac{aM\zeta_{\pm}[\sqrt{Mr}(r-3M)\pm2aM]^{1/2}}{\sqrt{r-2M}(Mr)^{3/4}}{\hat s}\right]+O({\hat s}^2)\nonumber\\
\frac1{\zeta}&=&\frac1{\zeta_{\pm} }+\frac32\frac{aM[\sqrt{Mr}(r-3M)\pm2aM]^{1/2}}{\sqrt{r-2M}(Mr)^{3/4}}{\hat s} +O({\hat s}^2)\ .
\end{eqnarray}
The total four momentum $P$ is given by (\ref{Ptot}) with
\begin{equation}
\frac{m_s}m=\frac{{\hat s}\gamma M^2}{r^2}\frac{r^2+a^2-a\nu\sqrt{\Delta }}{\sqrt{r-2M}\sqrt{r^3+a^2r+2a^2M}}\ , 
\end{equation}
and
\begin{equation}
\fl\quad \nu_p=\nu+{\mathcal N}^{(CP)}_p {\hat s}+O({\hat s}^2)\ , \quad 
{\mathcal N}^{(CP)}_p=\frac{M^2}{\gamma_{\pm}}\frac{r^2+a^2-a\sqrt{\Delta }\nu_{\pm}}{r^2\sqrt{r-2M}\sqrt{r^3+a^2r+2a^2M}}\ .
\end{equation}
The corresponding angular velocity $\zeta_p$ and its reciprocal are
\begin{eqnarray}
\label{zetapCP}
\zeta_p&=& \zeta+\frac{{\mathcal N}^{(CP)}_p r\sqrt{\Delta }}{r^3+a^2r+2a^2M}{\hat s}+O({\hat s}^2)\nonumber\\
\frac1{\zeta_p}&=& \frac{1}{\zeta}-\frac{{\mathcal N}^{(CP)}_p}{\zeta_{\pm}^2}\frac{r\sqrt{\Delta }}{r^3+a^2r+2a^2M}{\hat s}+O({\hat s}^2)\ .
\end{eqnarray}

To first order in the spacetime rotational parameter $a$ and neglecting terms of the order of $a\hat s$, the linear velocity (\ref{solCPexpnu}) and the reciprocal of the corresponding angular velocity (\ref{zetaCP}) are given by
\beq
\nu\simeq \pm\nu_K-3a\nu_K\zeta_K \ , \qquad 
\frac1{\zeta}\simeq  \frac1{\zeta_{\pm} }\ , 
\eeq
where the Keplerian quantities 
\beq\fl\quad 
\label{schwgeos}
\nu_K=\left[\frac{M}{r-2M}\right]^{1/2}, \qquad \zeta_K=\left(\frac{M}{r^3}\right)^{1/2}\ , \qquad \gamma_K=\left[\frac{r-2M}{r-3M}\right]^{1/2}\ 
\eeq
refer to timelike circular geodesics on the equatorial plane of the Schwarzschild spacetime.

\subsection{ The Pirani (P) supplementary conditions}

\begin{figure} 
\typeout{*** EPS figure 3}
\begin{center}
$\begin{array}{cccc}
\includegraphics[scale=0.35]{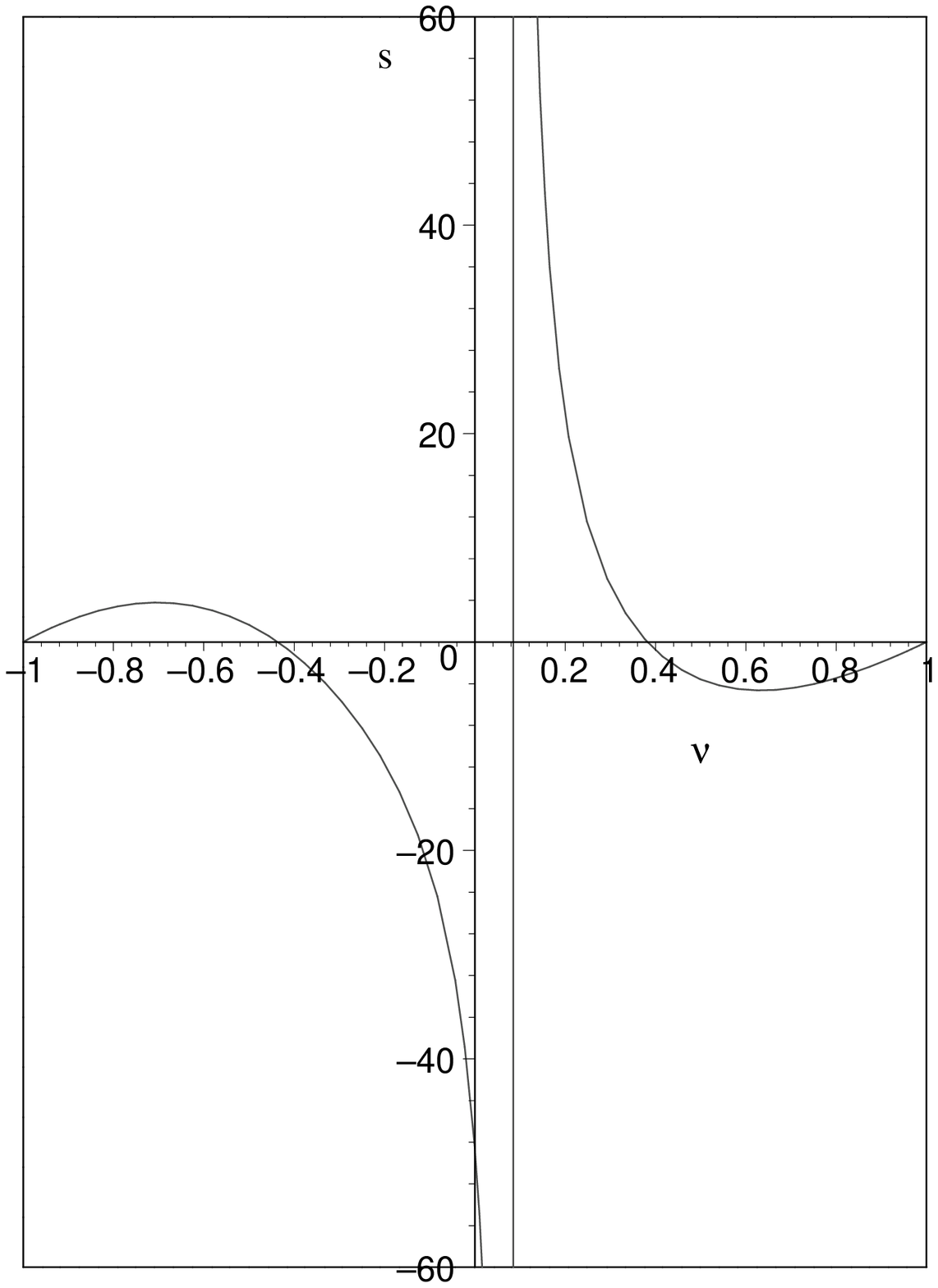}&\qquad
\includegraphics[scale=0.35]{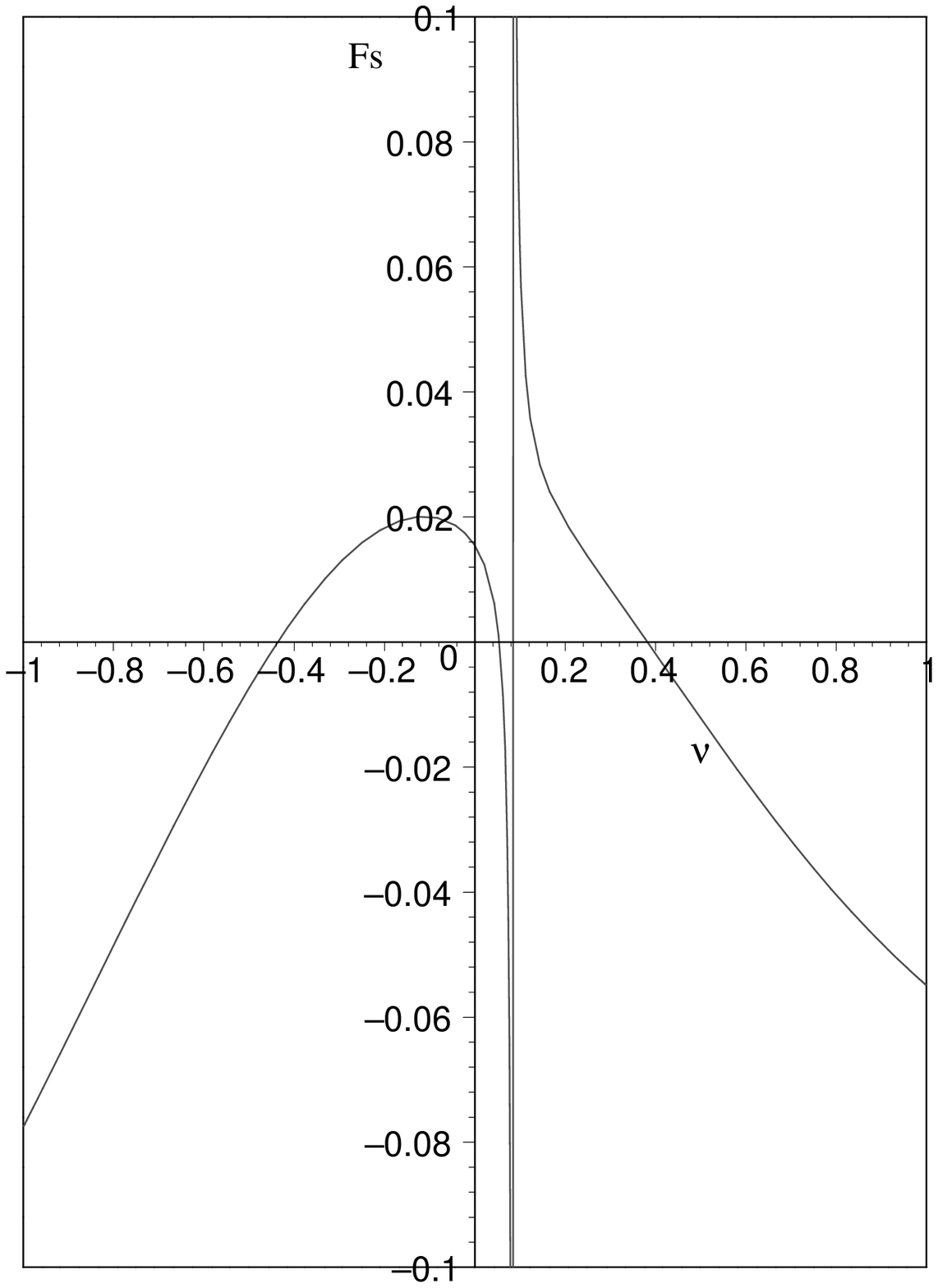}&\\[.2cm]
\qquad\mbox{(a)} &\qquad \mbox{(b)}&\\[.6cm]
\includegraphics[scale=0.35]{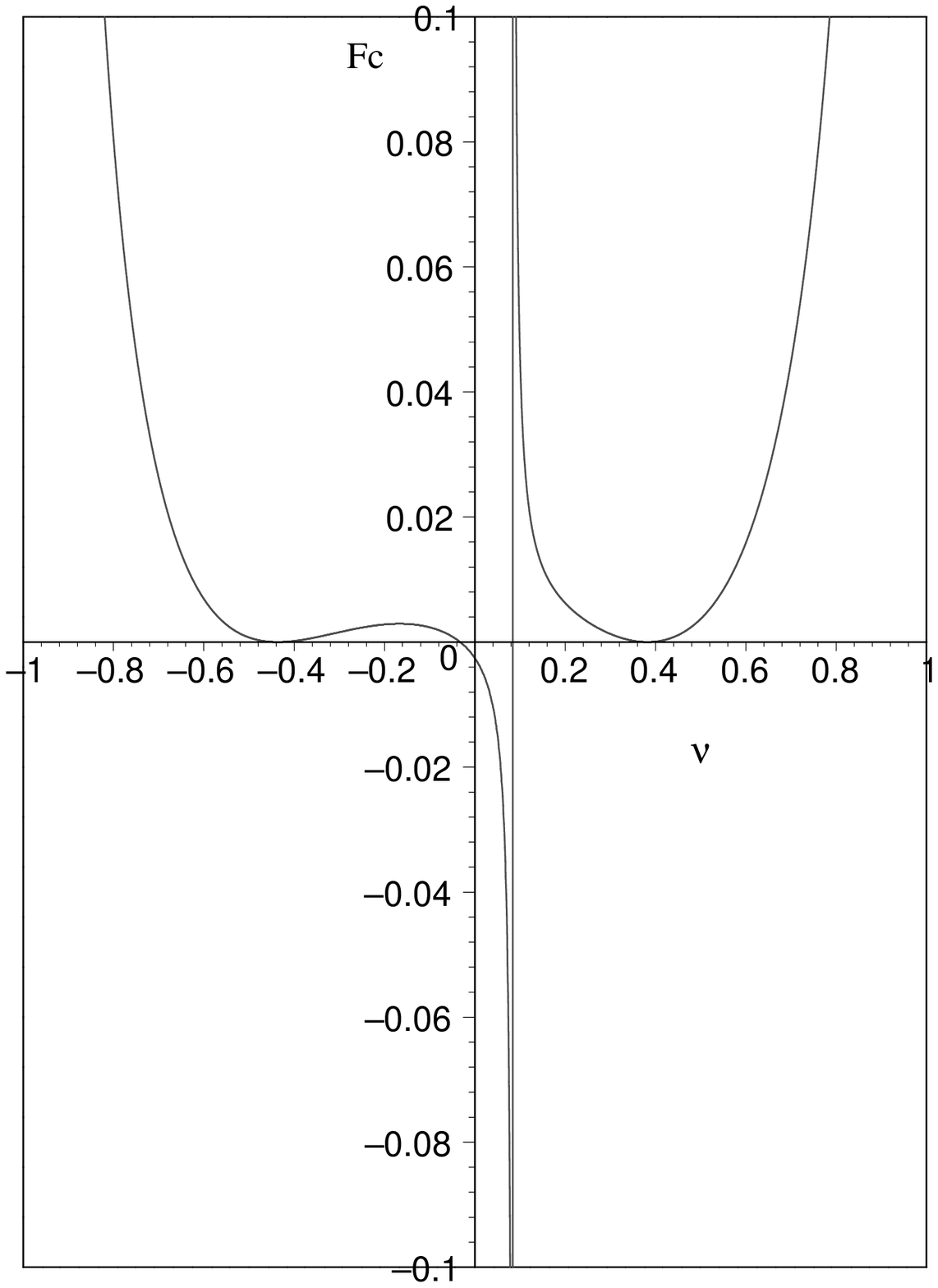}&\qquad
\includegraphics[scale=0.35]{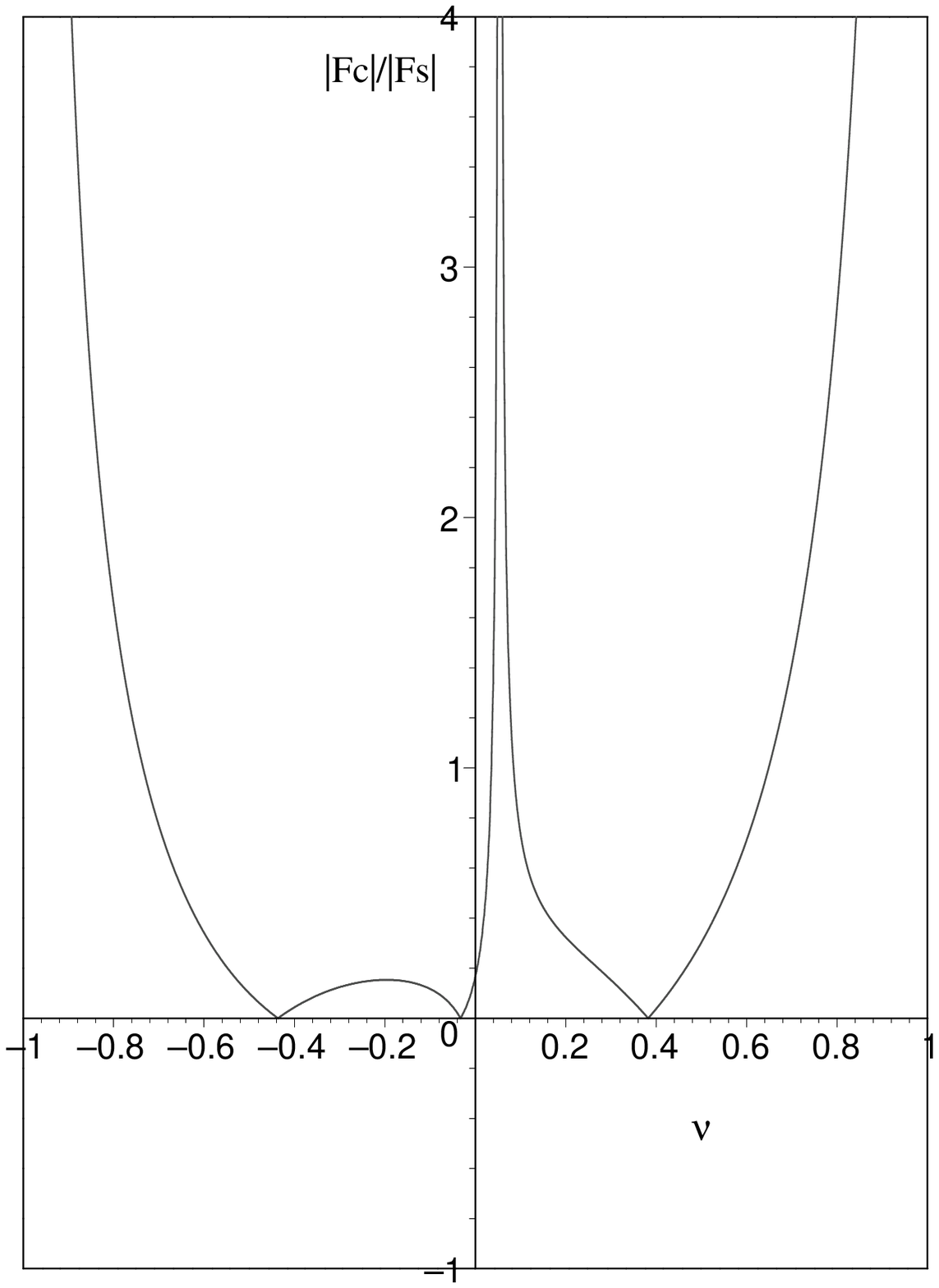}\\[.4cm]
\qquad\mbox{(c)}& \qquad\mbox{(d)}
\end{array}
$\\
\end{center}
\caption{In the case of P supplementary conditions, the spin parameter ${\hat s}$ is plotted in Fig. (a) as a function 
of the linear velocity $\nu$, for $M=1$, $a=0.5$ and $r=8$.
The corresponding behaviours of the spin and centrifugal forces and their ratio are shown in 
Fig. (b) - (d) respectively.
With this choice of parameters, the spin $\hat s$ diverges at $\nu\approx 0.084$ and the same is true for both the force due to the spin and the centrifugal force. 
Moreover, as the force due to the spin vanishes also at $\nu \approx 0.054$ the ratio between the centrifugal force and the spin force diverges at this value too. 
}  
\label{fig:3}
\end{figure}

The P supplementary conditions are given by (\ref{SCgen}) with $\tilde \nu=\nu$.  
The behaviour of the spin parameter ${\hat s}$ as a function of $\nu$ is shown in Fig. \ref{fig:3} (a), for a fixed values of $r$, $a$ and $M$. 
Comparisons between the centrifugal and spin forces are shown in Fig. \ref{fig:3} (b) - (d). Contrary to what happens in the CP case, the divergence of the spin $\hat s$
and the compensating vanishing of the spin force ($Fs$ in the plot)
now occur at positive values of $\nu$ (corotating orbits).

In the limit of small ${\hat s}$, Eq.~(\ref{ssolgen}) gives
\begin{eqnarray}
\label{solPexpnu}
\fl \nu&=& \nu_{\pm}+{\mathcal N}^{(P)}{\hat s}+O({\hat s}^2)\ , \\
\fl {\mathcal N}^{(P)}&=&\frac32\frac{M(r^3+a^2r+2a^2M)\{ar(r-5M)\pm\sqrt{Mr}[2a^2-r(r-3M)]\}^2}{r^2\sqrt{\Delta } \sqrt{Mr}(a\pm r\sqrt{r/M})^2}\cdot \nonumber\\
\fl\quad &&\cdot\{\sqrt{Mr}[4a^2(r-4M)-r(r-3M)^2]\pm a[4a^2M+r(r-3M)(r-7M)]\}^{-1}\ . \nonumber
\end{eqnarray}
The corresponding angular velocity $\zeta$ and its reciprocal are
\begin{eqnarray}
\label{zetaP}
\zeta&=& \zeta_{\pm} +\frac{{\mathcal N}^{(P)}r\sqrt{\Delta }}{r^3+a^2r+2a^2M}{\hat s}+O({\hat s}^2)\nonumber\\
\frac1{\zeta}&=&\frac1{\zeta_{\pm} }-\frac{{\mathcal N}^{(P)}}{\zeta_{\pm}^2}\frac{r\sqrt{\Delta }}{r^3+a^2r+2a^2M}{\hat s} +O({\hat s}^2)\ .
\end{eqnarray}
The total four momentum $P$ is given by (\ref{Ptot}) with
\begin{equation}
\frac{m_s}m=-M\kappa {\hat s}\ , 
\end{equation}
and
\begin{equation}
\nu_p=\nu+{\mathcal N}^{(P)}_p {\hat s}+O({\hat s}^2)\ , \qquad 
{\mathcal N}^{(P)}_p=-\frac{M\kappa}{\gamma_{\pm}^2}\ .
\end{equation}
The corresponding angular velocity $\zeta_p$ and its reciprocal are
\begin{eqnarray}
\label{zetapP}
\zeta_p&=& \zeta+\frac{{\mathcal N}^{(P)}_p r\sqrt{\Delta }}{r^3+a^2r+2a^2M}{\hat s}+O({\hat s}^2)\nonumber\\
\frac1{\zeta_p}&=& \frac{1}{\zeta}-\frac{{\mathcal N}^{(P)}_p}{\zeta_{\pm}^2}\frac{r\sqrt{\Delta }}{r^3+a^2r+2a^2M}{\hat s}+O({\hat s}^2)\ .
\end{eqnarray}

To first order in $a$ and neglecting also terms like $a\hat s$, the linear velocity (\ref{solPexpnu}) and the reciprocal of the corresponding angular velocity (\ref{zetaP}) become
\begin{eqnarray}
\label{nuzetaexpP}
\nu&\simeq&\pm\nu_K-3\nu_K\zeta_K\left[a+\frac{M{\hat s}}{2}\right]\nonumber\\ 
\frac1{\zeta}&\simeq&\frac1{\zeta_{\pm} }+\frac32 M{\hat s}\ . 
\end{eqnarray}

\subsection{ The Tulczyjew (T) supplementary conditions}

The T supplementary conditions are given by (\ref{SCgen}) with $\tilde \nu=\nu_p$.  
Recalling its definition (\ref{msdef2}), $m_s$ becomes
\begin{equation}
\frac{m_s}m=-M{\hat s}\gamma\gamma_p[-\nu_p (\tau_1+\kappa \nu )+(\nu \tau_1+\kappa )]\ ,
\end{equation}
and using (\ref{Ptot}) for $\nu_p$, we obtain 
\beq
\label{sfromms}
{\hat s}=\frac{1}{M\gamma\gamma_p}\frac{\nu-\nu_p}{(1-\nu\nu_p)[-\nu_p (\tau_1+\kappa \nu )+(\nu \tau_1+\kappa )]}\ , 
\eeq
which must be considered together with Eq.~(\ref{ssolgen}); of course, the case ${\hat s}=0$ (absence of spin) is only compatible with geodesic motion: $\nu\equiv\nu_p=\nu_{\pm}$.
By eliminating ${\hat s}$ from Eqs.~(\ref{ssolgen}) and (\ref{sfromms}) and solving with respect to $\nu_p$, we obtain
\begin{eqnarray}
\fl\quad \nu_p^{(\pm)}&=&\frac12\frac{(2\kappa \tau_1+A)\nu^2+[2(\kappa^2+\tau_1^2)+M/r^3]\nu+2\kappa \tau_1-A\pm\sqrt{\Psi}}{\kappa^2\nu^2+(2\kappa \tau_1+A)\nu+\tau_1^2+C}\label{nupsol1} \\ 
\fl\quad \Psi&=&[A^2+4\kappa (\kappa B+\tau_1 A)]\nu^4+[8A\kappa^2+2(B+C)(2\kappa \tau_1+A)]\nu^3\nonumber\\
\fl\quad &&+[4\kappa^2M/r^3+(B+C)^2+2A^2]\nu^2-[8A\kappa^2+2(B+C)(2\kappa \tau_1 -A)]\nu\nonumber\\
\fl\quad &&-A(4\kappa \tau_1-A)-4C\kappa^2\ 
\label{psi}.
\end{eqnarray}
Now, by substituting $\nu_p=\nu_p^{(\pm)}$ for instance into Eq.~(\ref{sfromms}), we obtain the main relation between ${\hat s}$ and $\nu$.

The reality condition of  (\ref{nupsol1}) requires that $\nu$ takes values outside the interval $({\bar \nu}_1,{\bar \nu}_2)$, 
with ${\bar \nu}_{1, 2}$ listed in Table \ref{tab:1} for some values of $a$ ($r, M$ fixed);
the timelike condition for $|\nu_p| <1$ is satisfied for all value of $\nu$ outside the same interval.


\begin{table}
\begin{center}
\begin{tabular}{|c|c|c|}
\hline
\rule{0pt}{3ex}$a$ & ${\bar \nu}_1$ & ${\bar \nu}_2$ \\ 
\hline
\rule{0pt}{2ex}$0$ & $-0.2970$ & $0.2970$ \\
\rule{0pt}{2ex}$0.2$ & $-0.2955$ & $0.2979$ \\
\rule{0pt}{2ex}$0.4$ & $-0.2933$ & $0.2984$ \\
\rule{0pt}{2ex}$0.6$ & $-0.2902$ & $0.2986$ \\
\rule{0pt}{2ex}$0.8$ & $-0.2861$ & $0.2984$ \\
\rule{0pt}{2ex}$1$ & $-0.2808$ & $0.2981$ \\
\hline
\end{tabular}
\end{center}
\caption{The limiting values of $\nu$ are listed for some values of the black hole rotational parameter $a$ and a fixed radial distance $r=8$ (M=1).}
\label{tab:1}
\end{table}

\begin{figure} 
\typeout{*** EPS figure 4}
\begin{center}
$\begin{array}{cccc}
\includegraphics[scale=0.35]{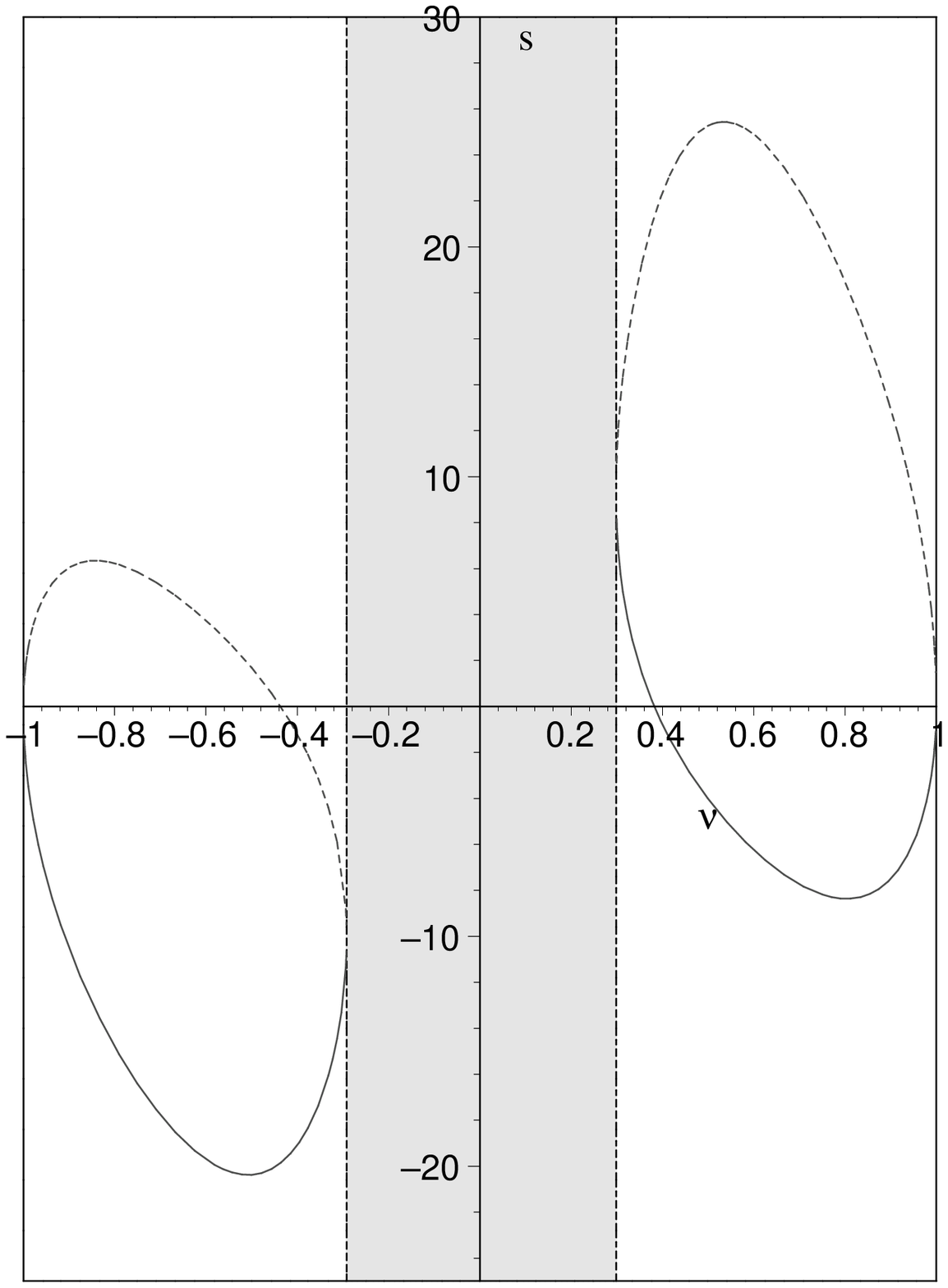}&\qquad
\includegraphics[scale=0.35]{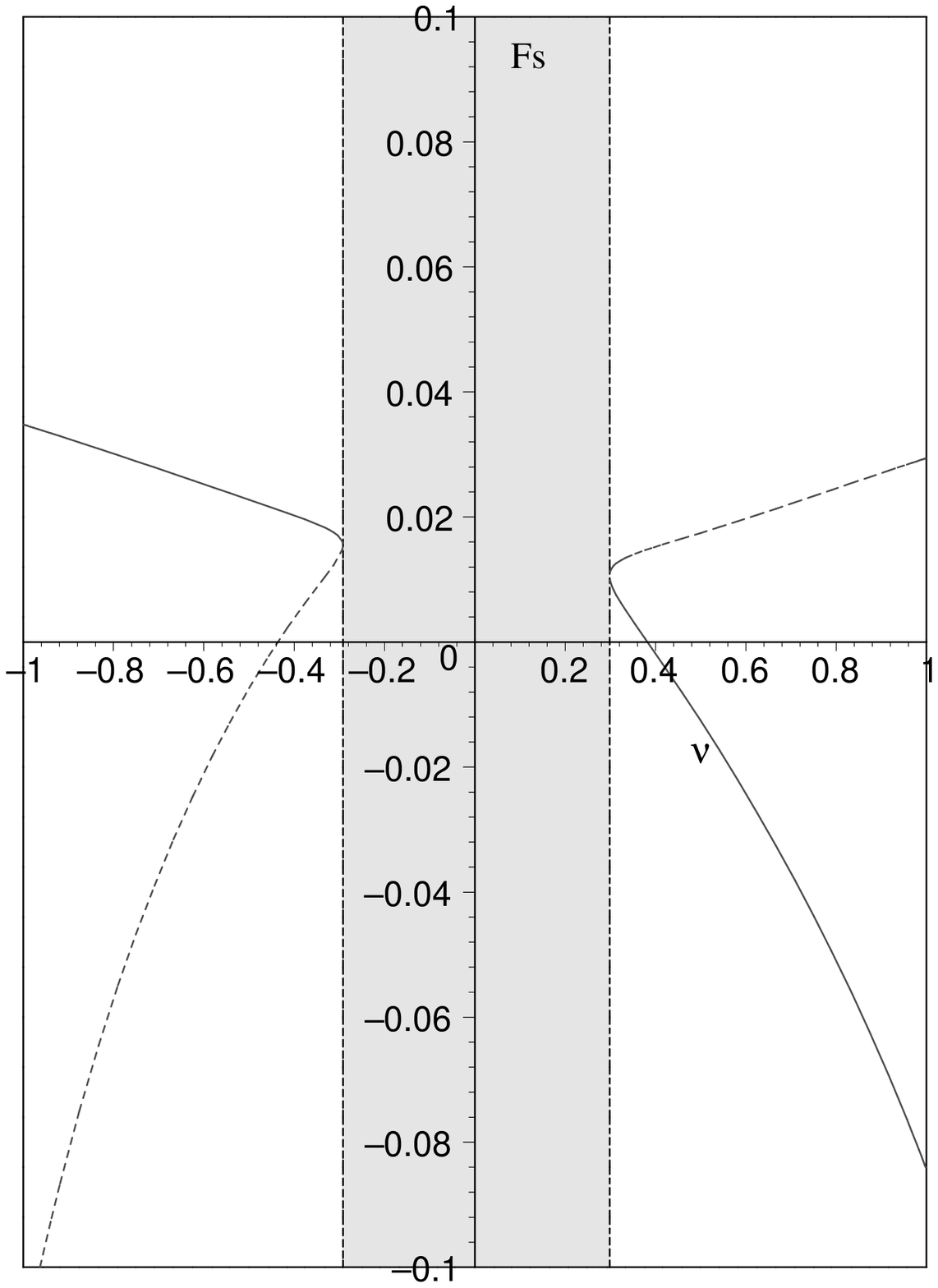}&\\[.2cm]
\qquad\mbox{(a)} &\qquad \mbox{(b)}&\\[.6cm]
\includegraphics[scale=0.35]{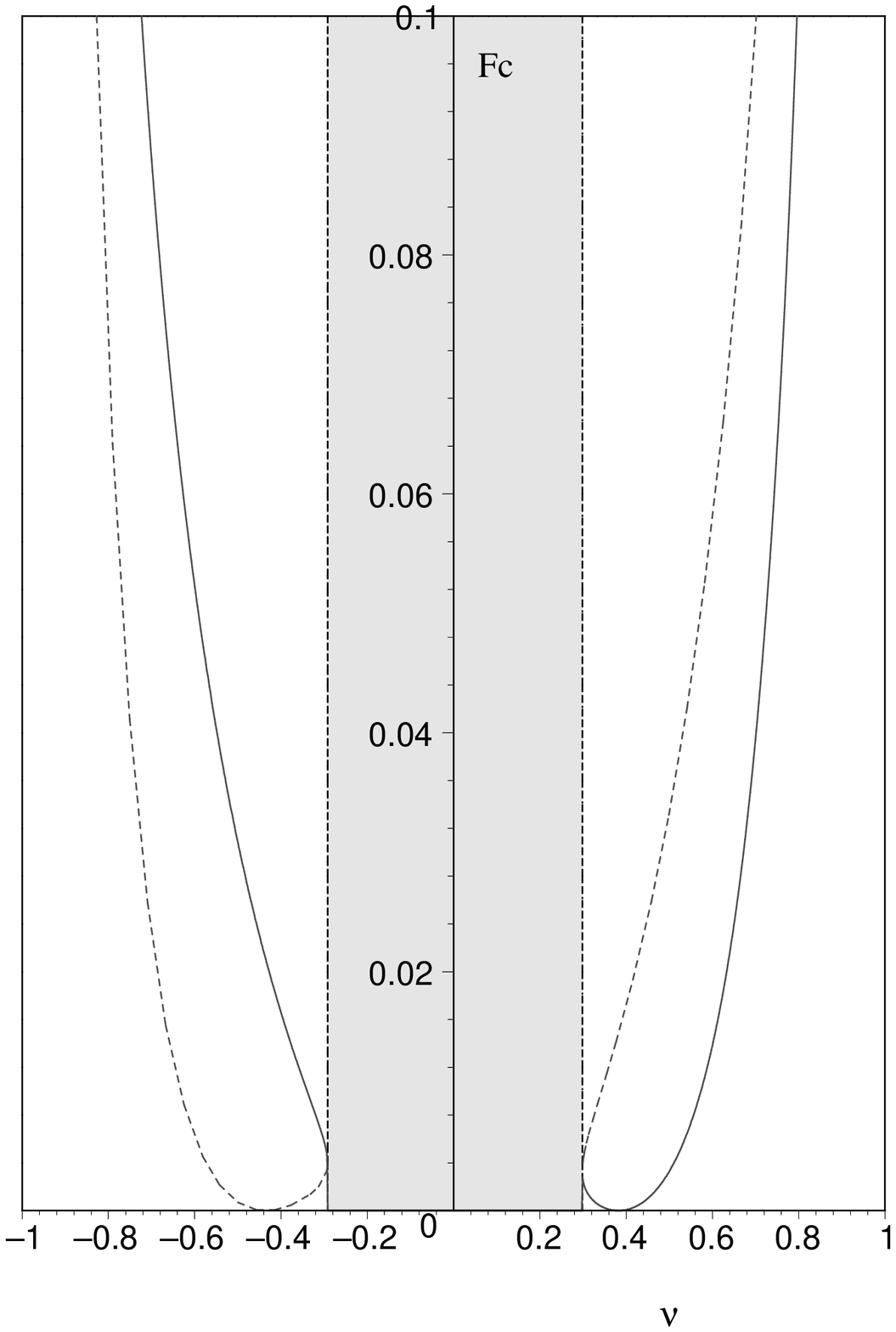}&\qquad
\includegraphics[scale=0.35]{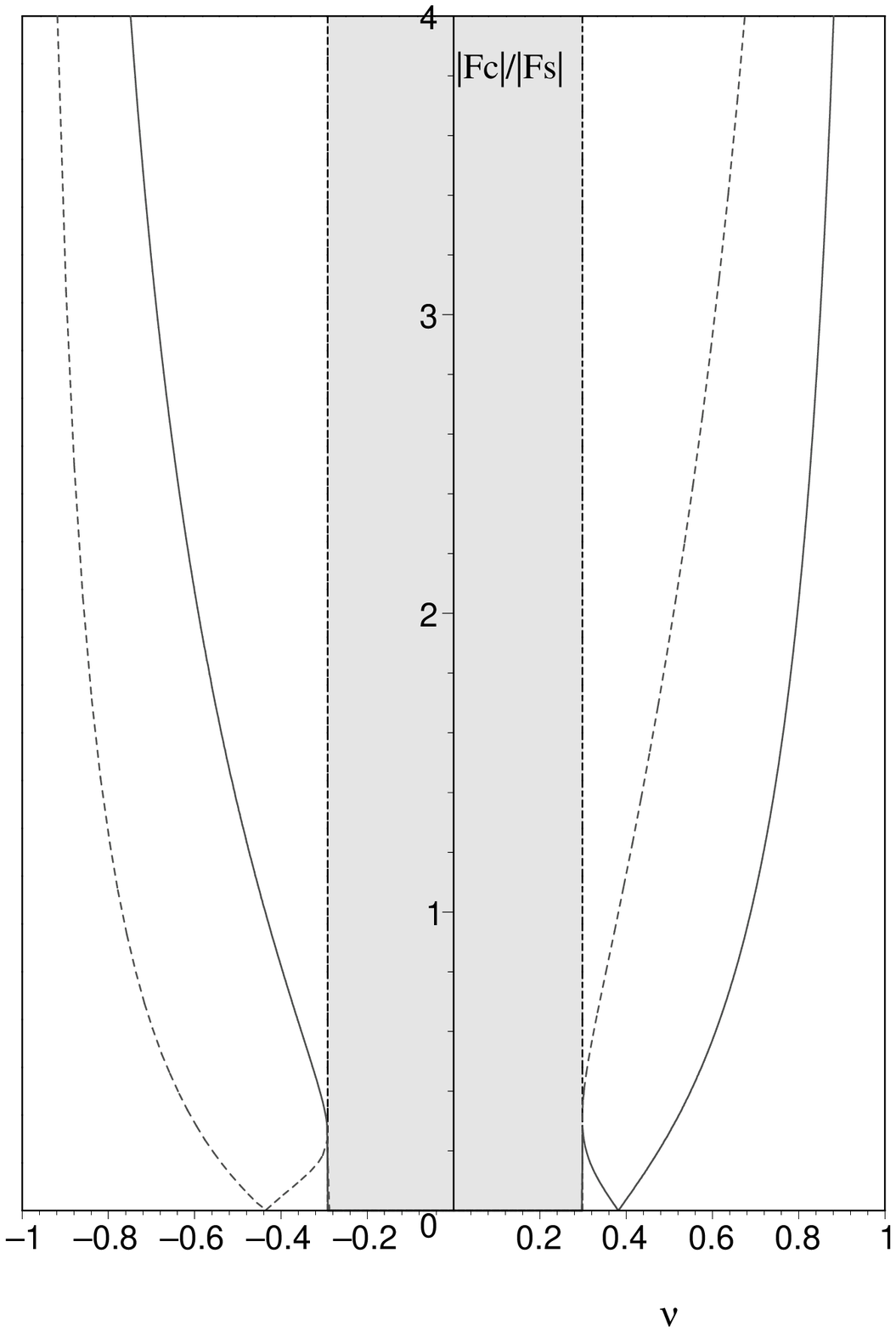}\\[.4cm]
\qquad\mbox{(c)}& \qquad\mbox{(d)}
\end{array}
$\\
\end{center}
\caption{In the case of T supplementary conditions, the spin parameter ${\hat s}$ is plotted in Fig. (a) as a function 
of the linear velocity $\nu$, for $r=8$, $M=1$, $a=0.5$ (and so ${\bar \nu}_{1}\approx-0.292$, ${\bar \nu}_{2}\approx 0.299$; the shaded region contains the forbidden values of $\nu$). The corresponding behaviours of the spin and centrifugal forces and their ratio are shown in 
Fig. (b) - (d) respectively. In Fig. (a) $\hat s(\bar \nu_1)=-10.065$, $\hat s(\bar \nu_2)=9.360$; in Fig. (b) $Fs(\bar \nu_1)=0.015$, $Fs(\bar \nu_2)=0.011$; in Fig. (c) $Fc(\bar \nu_1)=0.004$, $Fc(\bar \nu_2)=0.003$; in Fig. (d) $|Fc|/|Fs|(\bar \nu_1)=0.257$, $|Fc|/|Fs|(\bar \nu_2)=0.310$. 
The various curves have two branches corresponding to $\nu=\nu_p^{(+)}$ (solid) and to $\nu=\nu_p^{(-)}$ (dashed).
}  
\label{fig:4}
\end{figure}

The behaviour of the spin parameter ${\hat s}$ as a function of $\nu$ is shown in Fig. \ref{fig:4} (a).
Comparisons between the centrifugal and spin forces are shown in Fig. \ref{fig:4} (b) - (d).
The behaviour shown in those figures is qualitatively similar to what can be deduced in the Schwarzschild case; in particular we find now that no compensating vanishing of the spin force appears. 
This result may be used to observationally discriminate among the supplementary conditions.

To first order in $\hat s$ we have
\begin{eqnarray}
\label{solTexpnu}
\nu=\nu_{\pm}+{\mathcal N}^{(T)}{\hat s}+O({\hat s}^2)\ , \qquad
{\mathcal N}^{(T)}\equiv{\mathcal N}^{(P)}\ ; 
\end{eqnarray}
therefore, the angular velocity $\zeta$ and its reciprocal coincide with the corresponding ones derived in the case of P supplementary conditions (see Eq.~(\ref{zetaP})).
From the preceding approximate solution for $\nu$ we also have that 
\begin{equation}
\nu_p^{(\pm)}= \nu+O({\hat s}^2)\ , 
\end{equation}
and the total four momentum $P$ is given by (\ref{Ptot}) with $\nu_p=\nu_p^{(\pm)}$.
The corresponding angular velocity $\zeta_p$ and its reciprocal are
\begin{eqnarray}
\label{zetapT}
\zeta_p&=& \zeta+\frac{{\mathcal N}^{(T)} r\sqrt{\Delta }}{r^3+a^2r+2a^2M}{\hat s}+O({\hat s}^2)\nonumber\\
\frac1{\zeta_p}&=& \frac{1}{\zeta}-\frac{{\mathcal N}^{(T)}}{\zeta_{\pm}^2}\frac{r\sqrt{\Delta }}{r^3+a^2r+2a^2M}{\hat s}+O({\hat s}^2)\ .
\end{eqnarray}
To first order in $a$, the linear velocity (\ref{solTexpnu}) and the reciprocal of the angular velocity become equal to that derived for P supplementary conditions (see Eq.~(\ref{nuzetaexpP})).

\section{Clock-effect by spinning test particles}

In the case of geodesic spinless test particle, a static observer measures a gravitomagnetic time delay
between a pair of oppositely rotating timelike geodesics at a given radius given by 
\beq
\label{Dtnospin}
\Delta t_{(+,-)}=2\pi \left(\frac1{\zeta_+}+\frac1{\zeta_-}\right)=4\pi a\ .
\eeq
In the case of the Earth $a_\oplus\approx 3.4\times 10^2$ cm hence a clock effect amounts to a surprisingly high value of $\approx 10^{-7}$.

In the Schwarzschild case such an effect vanishes but if we consider spinning test particles  then, as shown in 
\cite{bdfg1}, the time delay is nonzero and  can be measured by using co/counter-rotating spin-up/spin-down particles.

In the Kerr case, the spin of the particle is responsible of a further contribution to (\ref{Dtnospin}).
In fact,  in all the cases examined above, circularly rotating spinning test particles, to the first order in the spin parameter $\hat s$, $a$ and neglecting terms as $\hat s a$, have orbits close to the geodesics (as expected), with
\beq
\label{clock}
\frac{1}{\zeta_{(SC,\pm,\pm)}}=\frac{1}{\zeta_{\pm} } \pm M|{\hat s}|{\mathcal J}^{SC}\ ,
\eeq 
 with ${\mathcal J}^{CP}=0$, ${\mathcal J}^{P}={\mathcal J}^{T}=3/2$.
Eq.~(\ref{clock}) defines such orbits corresponding to different supplementary conditions  with the various signs corresponding to co/counter-rotating orbits with positive/negative spin direction along the $z$ axis. For instance, the notation $\zeta_{(P,+,-)}$ indicates the angular velocity of $U$, derived under the choice of Pirani's supplementary conditions and corresponding to a co-rotating orbit $(+)$ with spin-down $(-)$ alignment, etc.
Therefore one can study the difference in the arrival times after one complete revolution with respect to a local static observer:
\begin{eqnarray}
\Delta t_{(+,+;-,+)}&=& 2\pi \left(\frac{1}{\zeta_{(SC,+,+)}}+\frac{1}{\zeta_{(SC,-,+)}}\right)=
4 \pi \left(a+M|{\hat s}| {\mathcal J}^{SC}\right)\nonumber\\
\Delta t_{(+,+;-,-)}&=&\Delta t_{(+,-;-,+)}=4 \pi a\nonumber\\
\Delta t_{(+,-;-,-)}&=&4 \pi \left(a-M|{\hat s}|{\mathcal J}^{SC}\right)\ .
\end{eqnarray}
In the latter case, it is easy to see that if $a=\frac32 M|\hat s|$ the clock-effect can be made  vanishing \cite{faruq}; this feature which can be
directly measured. More interesting is the case of corotating spin-up against counter-rotating spin-down or alternatively co-rotating spin-down against counter-rotating 
spin-up. Now the compensating effect makes the spin contribution to the clock effect equal to zero; this case therefore appears indistinguishable from that of  spinless particles.

\section{Conclusions}

Spinning test particles in (equatorial) circular motion around the kerr black hole have been discussed in the framework of the Mathisson-Papapetrou approach supplemented by standard conditions, generalizing a previous discussion carried out in the case of the Schwarzschild spacetime.
In general, spinning test particles can maintain a circular orbit rotating with an angular velocity which is spin-dependent.
We notice that in the limit of small spin, the orbit of the particle is close to a circular geodesic and
the difference in the angular velocities with respect to the geodesic value might be of arbitrary sign, corresponding to the two spin-up and spin-down orientations along the $z$-axis. It is well known that for spinless particles the difference in the arrival times  of co-rotating and counter-rotating circular geodesics as measured by a static observer, namely the gravitomagnetic clock effect,  is non-zero and proportional to the spacetime rotation 
itself. In the presence of spin, the clock-effect is modified; in the limit of small spin we find as already shown in \cite{faruq} that this effect may be
 amplified or inhibited and even suppressed.
This result  can be tested by experiments.

\section*{Acknowledgments}
We are grateful to Prof. B. Mashhoon and Prof. R.T. Jantzen for a critical reading of the manuscript.
Part of this work was done when one of us (F. de F.) was at the Instituto Venezolano de Investigaciones Cientificas (IVIC)
nearby Caracas. The director of that institution is thanked for his warm hospitality and the Consiglio Nazionale delle Ricerche (CNR) of Italy is thanked for support. 
D.B. acknowledges discussion with Dr. G. Organtini too.

\end{document}